\documentclass[preprint,11pt]{elsarticle} 
\makeatletter
\def\ps@pprintTitle{%
	\let\@oddhead\@empty
	\let\@evenhead\@empty
	\def\@oddfoot{}%
	\let\@evenfoot\@oddfoot}
\makeatother
\usepackage[left=15mm,right = 15mm,top = 15mm,bottom = 15mm]{geometry}
\usepackage[titletoc]{appendix}
\usepackage{CJKutf8}

\usepackage[table]{xcolor}
\usepackage{float}
\usepackage{subcaption}
\usepackage{amssymb,amsmath,amsthm,graphics,psfrag,graphicx,color,fancyhdr}
\usepackage{algorithmic}
\usepackage[color=yellow]{todonotes}

\usepackage{appendix}
\usepackage{verbatim}
\usepackage{blindtext}
\usepackage{hyperref}
\usepackage{cleveref}
\usepackage{stmaryrd}
\usepackage{soul}
\usepackage{sidecap}
\usepackage{tikz}
\usepackage{rotating}
\usepackage{booktabs}
\usepackage{comment}
\usepackage{multirow}
\usepackage{changepage}
\usepackage{float}
\usepackage{wrapfig}
\usepackage{stmaryrd}
\usepackage{pgfplots}
\usepackage{framed}
\usepackage{float}

\usepgfplotslibrary{fillbetween}

\crefformat{appendix}{#2#1#3}

\definecolor{lightblue}{rgb}{0.32,0.45,0.90}
\definecolor{lightgreen}{rgb}{0.42,0.7,0.40}
\numberwithin{equation}{section}
\numberwithin{figure}{section}
\hypersetup{colorlinks=true,linkcolor = blue,citecolor=blue}
\usetikzlibrary{shapes,arrows}
\usetikzlibrary{decorations.markings}
\numberwithin{figure}{section}

\def\b{\boldsymbol}

\makeatletter
\newcommand{\vast}{\bBigg@{4}}
\newcommand{\Vast}{\bBigg@{5}}

\makeatother\def\b{\boldsymbol}

\colorlet{cgray}{gray!20!white}

\theoremstyle{definition}
\newtheorem{definition}{Definition}[section]

\newtheorem{method}{Method}[section]

\theoremstyle{remark}

\tikzset{every label/.style={font=\footnotesize,inner sep=1pt}}

\allowdisplaybreaks

\pgfplotsset{compat=1.18} 

\begin{document}
\begin{frontmatter}
\title{Monotone conservative strategies in data assimilation}
\author[inst4]{J. Woodfield}

\affiliation[inst4]{organization={Department of Mathematics, Imperial College London},
addressline={South Kensington Campus}, 
city={London},
postcode={SW7 2AZ}, 
country={United Kingdom}}

\begin{abstract}
This paper studies whether numerically preserving monotonic properties can offer modelling advantages in data assimilation, particularly when the signal or data is a realization of a stochastic partial differential equation (SPDE) or partial differential equation (PDE) with a monotonic property. 
We investigate the combination of stochastic Strong Stability Preserving (SSP) time-stepping, nonlinear solving strategies and data assimilation. Experimental results indicate that a particle filter whose ensemble members are solved monotonically can increase forecast skill when the reference data (not necessarily observations) also has a monotone property. Additionally, more advanced techniques used to avoid the degeneracy of the filter (tempering-jittering) are shown to be compatible with a conservative monotone approach. 
\end{abstract}
\end{frontmatter}
\tableofcontents

\section{Introduction}

\subsection{Motivation}

Stochastic parametrisations of various types \cite{buizza1999stochastic,holm2015variational,bismut1976linear,leutbecher2017stochastic,hagelin2017met,sanchez2016improved,mccabe2016representing} have shown to be effective in quantifying different types of uncertainty, reducing model error and increasing forecast skill and reliability. There is increased interest in producing ensemble members as the realisation of SPDE's for the use in uncertainty quantification see for example \cite{breit2024discontinuous,cotter2018modelling,lobbe2024bayesian,hu2023variational,breit2016stochastic} where stochastic versions of the Euler, Quasi-geostrophic, Shallow-Water, Primative, and Navier-Stokes equations have been studied and proposed. Recently, deterministic notions of strong stability preserving time integration outlined in (\cite{shu1988efficient,spijker1983contractivity,higueras2004strong,bolley1978conservation,ketcheson2004algebraic,kraaijevanger1991contractivity,higueras2005representations}) were shown to translate into the stochastic context, allowing non-linearly stable numerical solutions to SPDE's and SDE's \cite{woodfield2024strong}. Where it was hypothesised that a monotonic approach
may offer modelling advantages in data assimilation, particularly when the signal or data is a realization of an SPDE or PDE with a monotonicity property.

This paper demonstrates (through a simple proof of concept experiment) practical advantages in maintaining monotonic properties in the ensemble, whilst performing sequential data assimilation for the nonlinear-advection equation with a particle filter. Then in anticipation of higher dimensional systems it is shown jittering in the particle filter can be performed monotonically, and potentially with reduced cost than rerunning the forward model.


\subsection{Outline}

In \cref{sec:Background: SSP time integration}, deterministic and stochastic strong stability preserving time integration are introduced. In \cref{sec:minimising diffusion} two numerical approaches are proposed for a stochastic constant coefficient advection equation with pathwise total variation diminishing bounds, and it is justified that by combining stochastic and deterministic velocity fields into a single velocity one minimises numerical diffusion when using an upwinding approach. In \cref{sec:Equations and monotone solving strategy} we introduce a nonlinear compressible stochastic transport equation and a positivity-preserving solving strategy. 

In \cref{sec:Example 2a: The Twin Experiment} we produce ensemble solutions to a one dimensional stochastic compressible advection equation with and without a particle filter and compute ensemble forecast metrics when the data is a realisation of the monotone forward model. 
In \cref{sec:Example 2b: One-dimensional slope limiters} we perform the same experiment, however, we no longer use idealised data but generate reference data from a higher resolution PDE with a similar monotonic property. We show an increase in continuous ranked probability score as a function of lead time when using a monotonic solving strategy, indicating potential modelling advantages in creating ensembles with monotonic properties.

In \cref{sec:degeneracy in the particle filter} two preliminary experiments are performed as to indicate, how one could jitter monotonically. The first experiment performed one step of monotonic jittering (without tempering) with a Metropolis-Hastings accept-reject step for idealised data. The second experiment performed many stages of tempering with a cheaper version of frozen-in-time monotonic flux form drift-free advection jittering, in the context of non-idealised data. Similar CRPS and RMSE were observed to additive jittering but with positivity preservation.

\subsection{Background: SSP time integration}\label{sec:Background: SSP time integration}
As first remarked upon by Shu and Osher in 1988 \cite{shu1988efficient}, an explicit $s$-stage Runge-Kutta method can often be rearranged into a Shu-Osher representation, where the first substage is initialised at the old timestep $k^{0} = u^n$, to update the new solution $ k^{s}=u^{n+1}$, in the following structure
\begin{align}
k^{i} & =\sum_{j=0}^{i-1}\alpha_{i j}\left( k^{j} + \Delta t \frac{\beta_{i j}}{\alpha_{i j}} f(k^{j})\right), \quad i\in \lbrace 1, ... , s \rbrace \label{eq:shu-osher-representation}.
\end{align}
 The motivation behind the representation \cref{eq:shu-osher-representation}, is that the conditions 
\begin{align}
    \sum_{j=0}^{i-1}\alpha_{ij} = 1, \quad \alpha_{ij}\geq 0,\quad \beta_{ij}\geq 0, \quad \beta_{ij}=0\implies \alpha_{ij}=0.
\end{align}
are sufficient for \cref{eq:shu-osher-representation} to be a convex combination of Forward Euler schemes with modified timesteps. This convex combination representation allows contractive and nonlinear properties to be proven about the Runge-Kutta scheme provided one can establish the same property for the Forward Euler flow map. For example let $||\cdot||$ denote an arbitrary convex functional where it is assumed
\begin{align}
||\operatorname{FE}(u,\Delta t)|| = ||u + \Delta t f(u)||\leq ||u||,\quad \forall \Delta t \leq \Delta t_{FE}\label{eq:Forward Euler assumption}.
\end{align}
then by the representation \cref{eq:shu-osher-representation} the triangle inequality and the timestep condition 
\begin{align}
    \Delta t \frac{\beta_{i j}}{\alpha_{i j}} \leq \Delta t_{FE}, \quad \forall i, j.
\end{align}
one can derive the property,
\begin{align}
||k^{i}|| \leq \sum_{j=0}^{i-1}\alpha_{ij}|| k^{j} + \Delta t \frac{\beta_{i j}}{\alpha_{i j}} f(k^{j})|| \leq  \sum_{j=0}^{i-1}\alpha_{ij}||k^{j}|| \leq ||u^n||, \quad \forall i \in \lbrace 1,...,s \rbrace.
\end{align}

Motivating the radius of monotonicity or SSP coefficient of this explicit stochastic Runge-Kutta scheme
\begin{align}
\Delta t \leq C\Delta t_{FE},\quad 
C = \text{min}_{i,j}\frac{\beta_{ij}}{\alpha_{ij}}.
\end{align}

One should note that the coefficients $\alpha_{ij},\beta_{ij}$, are related but not uniquely specified by the Butcher tableau of the Runge-Kutta method, and extensive theoretical and computational research has been dedicated to the finding of optimal representations of Runge-Kutta methods with large radii of monotonicity. These ideas have recently been shown to extend to SDE's/SPDE's in \cite{woodfield2024strong}. The simplest setting for which the theory translates is for Stratonovich SDE's of the form
\begin{align}
dq + f(q) + \sum_{p=1}^{P} g_p(q)\circ d W^p = 0,
\end{align}
when stochastic explicit Runge-Kutta methods satisfying the order two butcher conditions admit the following Euler-Maruyama convex combination representation 
\begin{align}
k^{0} & = u^n, \\
k^{i} & =\sum_{j=0}^{i-1}\alpha_{i j}\left( k^{j} + \Delta t \frac{\beta_{i j}}{\alpha_{i j}} f(k^{j})+ \frac{\beta_{i j}}{\alpha_{i j}}  \sum^{P}_{p=1} g_p(k^j) \Delta S^p\right) , \quad i=1, \ldots, s , \label{eq:stochastic-shu-osher-representation}\\
u^{n+1} & =k^{s}.
\end{align}
The second-order butcher tableau conditions are required for Stratonovich convergence \cite{ruemelin1982numerical}. 
Here it is assumed $\Delta S^p$ are bounded such that one can prove the following Euler-Maruyama property, 
\begin{align}
||u + \Delta t  f(u)+ \sum^{P}_{p=1} g_p(u) \Delta S^p|| \leq || u ||, \quad \forall \Delta t \leq \tau_{EM}, \quad \forall \Delta S^p. \label{eq: SSP-EM property}
\end{align}
Then analogous to the deterministic setting, one establishes 
\begin{align}
    ||k^i||\leq ||u^n||, \quad \forall i \in \lbrace 1,...,s\rbrace, \quad \text{provided}\quad \Delta t\leq \min_{i,k}\left(\frac{\beta_{ij}}{\alpha_{ij}}\right) \tau_{EM},
\end{align}
for the stochastic Runge-Kutta method \cref{eq:stochastic-shu-osher-representation}. Strong convergence is not lost using the bounded increments in \cite{milstein2002numerical} defined as $\Delta \widetilde{W} = \sqrt{\Delta t} \Delta \widetilde{Z}_{\Delta t}$ where the $\Delta \widetilde{Z}_{\Delta t}$-r.v. is bounded by $A_{\Delta t}:=\sqrt{2 k|\ln {\Delta t}|}, k \geq 1$, and defined by
\begin{definition}
[Milstein-Tretyakov Bounded increments]\label{Milstein-Tretyakov Bounded normal increments} Given $\Delta Z\sim N(0,1)$ and $\Delta t \in \mathbb{R}^{>0}$, Milstein and Tretyakov \cite{milstein2004stochastic} define a symmetric bounded increment $\Delta \widetilde{Z}_{\Delta t}$, from the following random variable, 
 \begin{align}
\Delta \widetilde{Z}_{\Delta t}:=\left\{\begin{array}{c}
\Delta Z,\quad |\Delta Z| \leq A_{\Delta t}, \\
A_{\Delta t}, \quad \Delta Z>A_{\Delta t},\\
-A_{\Delta t}, \quad \Delta Z<-A_{\Delta t}.
\end{array}\right.
\label{eq:bounded increments}
\end{align}
\end{definition}
For details of convergence see \cite{milstein2002numerical,milstein2013numerical} additional details can be found in \cite{woodfield2024stochastic}.

Even with bounded increments one may be sceptical about proving \cref{eq: SSP-EM property} as $\Delta S$ can take both positive and negative values. For instance assuming the SSP property \cref{eq: SSP-EM property}, setting $f=0$, $P=1$, $g_1(u)\neq 0$, 
gives the scheme 
\begin{align}
u^{n+1} = u + g_1(u) \Delta S^1
\end{align}
and choosing $\Delta S^1 = \pm 1$, gives both $
||u+ g_1(u)||\leq ||u||$, $||u- g_1(u)||\leq ||u||$, implying $g_1(u) = 0$, a contradiction. Dissuading a notion of monotonicity of the Euler Maruyama scheme. This argument holds, and the implications are perhaps understated in \cite{woodfield2024strong}. However, one can circumnavigate this inconvenience using a well known technique originally introduced by Shu and Osher in \cite{shu1988efficient} to deal with negative coefficients in the Runge-Kutta method. One defines a new operator $\mathfrak{g}$, such that one can establish both
\begin{align}
|| u + g_1(u) \Delta S^1 || \leq ||u||, \quad \forall \Delta t \leq \tau_{EM}, \quad \forall \Delta S^1 \in [0,A],\\
|| u + \mathfrak{g}_1(u) \Delta S^1 || \leq ||u||, \quad \forall \Delta t \leq \tau_{EM}, \quad \forall \Delta S^1 \in [-A,0].
\end{align}
Then one adaptively chooses either $g$, $\mathfrak{g}$, based upon the sign of $\Delta S^1$, making the numerical method itself a stochastic process to ensure monotone properties. The classical example is changing an upwind discretisation with a negative timestep into a downwind bias discretisation with a positive timestep. 

In the next section, this is discussed more concretely and \cref{eq: SSP-EM property} is established in the total variation semi-norm for two separate discretisations of the stochastic advection equation under a constant translational noise. This is done primarily to indicate how to ensure the SSP property in the situation in which $f\neq 0$, and $P>1$, and discuss how one can minimise numerical diffusion by combining velocities.

\subsection{Background: Minimising diffusion}\label{sec:minimising diffusion}

Consider discretising the linear advection equation with constant translational noise
$$du + au_x dt + u_x dW = 0,\quad a>0,$$ with
a Euler Maruyama first-order upwind scheme 
\begin{align}
EM(u)_i = u_i - a \Delta t/\Delta x [u_{i}-u_{i-1}] - [u_{i}-u_{i-1}] (\Delta \widetilde{W})^+ / \Delta x -  [u_{i+1}-u_{i}] (\Delta \widetilde{W})^- / \Delta x  ,\label{eq:eulermaruyama1}
\end{align}
where $(\cdot)^+ := \max(\cdot ,0)$, $(\cdot)^- := \min(\cdot ,0)$. Based on the sign of the realisation of the bounded random variable $\Delta \widetilde{W}$, one can ensure that a monotone upwind scheme is chosen. Therefore, since $\left( a\Delta t/\Delta x + (\Delta \widetilde{W})^{+}/\Delta x \right)\geq 0$, $\left(-(\Delta \widetilde{W})^{-}/\Delta x\right)\geq 0$, under the timestep(CFL) condition $
0\leq a \Delta t/\Delta x + |\Delta \widetilde{W}|/\Delta x \leq 1$,
one can establish the Total Variation Diminishing (TVD) property
\begin{align}
|EM(u)|_{BV}:=\sum_{i}|EM(u)_{i}-EM(u)_{i-1}| &\leq \sum_{i} |u_{i} - u_{i-1}|\left( 1 - a \Delta t/\Delta x -(\Delta x)^{-1}\left((\Delta \widetilde{W})^+ +(\Delta \widetilde{W})^- \right)\right)\\
& + \sum_{i}|u_{i-1}-u_{i-2}| \left( a\Delta t/\Delta x + (\Delta \widetilde{W})^{+}/\Delta x \right)\\
&+ \sum_{i}|u_{i+1}-u_i|\left(-(\Delta \widetilde{W})^{-}/\Delta x\right) \leq |u|_{BV}.
\end{align}
Any strong stability preserving stochastic Runge-Kutta method capturing the Itô-Stratonovich correction (\cref{eq:stochastic-shu-osher-representation}) would converge to the Stratonovich SDE $du + au_x dt + u_x \circ dW = 0$, and inherit the total variation stability property
\begin{align}
|u^{n+1}|_{BV} \leq |u^n|_{BV}, \quad \forall n.
\end{align}
from the Euler-Maruyama scheme with bounded increments,
under the stochastic extension of the SSP theory \cite{woodfield2024strong}.

Whilst this is a TVD scheme, one can minimise some of the artificial diffusion, by combining the velocity and defining the following different numerical scheme
\begin{align}
EM(u)_i = u_i - \Delta t/\Delta x \left( a  + \Delta \widetilde{W}/\Delta t\right)^{+}[u_{i}-u_{i-1}] - \Delta t/\Delta x\left( a  + \Delta \widetilde{W}/\Delta t\right)^{-} [u_{i+1}-u_{i}]. \label{eq:eulermaruyama2} 
\end{align}
This different scheme also satisfies the TVD property, with a slightly different CFL condition
\begin{align}
\frac{\Delta t}{\Delta x}\left| \left(a +\Delta \widetilde{W} /\Delta t\right)\right| \leq 1 \implies |\operatorname{EM}(u)|_{BV} \leq |u|_{BV},
\end{align}
To see how diffusion is minimised using the second approach consider a first-order (monotone-TVD) upwind scheme of $u_x$ this has the following leading order truncation error (LOTE),
\begin{align}
u_x \approx \frac{u_i - u_{i-1}}{\Delta x} = u_x - \underbrace{1/2\Delta x u_{xx}}_{LOTE} + \mathcal{O}(\Delta x^2).
\end{align}
Then consider splitting the velocity as follows $u_x=(a-b)u_x$, where $b>0$, $1 = a-b$, $a>1$, using an upwinding-downwinding strategy, one calculates a larger leading order diffusion truncation error for all $b>0$,
\begin{align}
u_{x} &\approx a\frac{u_i - u_{i-1}}{\Delta x}   - b \frac{u_{i+1} - u_{i}}{\Delta x} \\
&\approx \frac{u_i - u_{i-1}}{\Delta x} - b \frac{u_{i+1}-2u_i +b u_{i-1}}{\Delta x} = u_x - \underbrace{\Delta x\left(1/2 + b\right) u_{xx}}_{LOTE} + \mathcal{O}(\Delta x^2).
\end{align}
Concluding that splitting a positive velocity into a positive and negative velocity when using an upwind approach leads to more diffusion than when leaving the velocity combined. Conversely, by combining positive and negative velocities we can minimise numerical diffusion, so that \cref{eq:eulermaruyama2} is less diffusive than \cref{eq:eulermaruyama1}. Combining positive and negative velocities when considering higher-order upwind bias operators, is also analogously expected to minimise (higher-order) diffusion. This motivates the numerical following strategy of the \emph{compressible} advection equation.

\subsection{Equations and monotone solving strategy}\label{sec:Equations and monotone solving strategy}

We consider the following stochastic compressible nonlinear transport equation 
\begin{align}
d q(t,x) + ( u(t,x) q(t,x))_x dt + \sum_{p=1}^{P}( \xi_{p}(x) q(t,x))_x\circ dW^{p}_t= 0,\quad q(0,x)=q_0(x) \label{eq:stochastic 1d transport problem}
\end{align}
where $P=16$, and the compressible vector fields are specified as follows
\begin{align}
u(x) =\frac{1}{20}\left(9+\sin(2\pi x)\right), \quad \xi_p(x) =  \frac{3}{25\pi^2}\frac{1}{p^2} \sin(2\pi p x), \quad p\in \lbrace 1,2,3,4,...,16\rbrace.
\end{align}
Subject to the initial condition
\begin{align}
q_0(x)=
\begin{cases}
\sin(4 \pi x) \quad &\text{where}\quad (x<0.25),\\
1  \quad &\text{where}\quad (0.5<x<0.8), \\
0 \quad & \text{else}.
\end{cases}
\end{align}

To solve \cref{eq:stochastic 1d transport problem} monotonically, we use
the noise increments in \cref{eq:bounded increments},
bounded by $A_{\Delta t} = \sqrt{2 |\log(\Delta t)|}$, and design a monotonic EM scheme for the Itô system
\begin{align}
d q(t,x) + ( u(t,x) q(t,x))_x dt + \sum_{p=1}^{P}( \xi_{p}(x) q(t,x))_x dW^{p}_t= 0,\quad q(0,x)=q_0(x) \label{eq:ito stochastic 1d transport problem}.
\end{align}

Then the stochastic SSP33 time-stepping method combines the $\operatorname{EM}$ scheme in a convex combination representation as follows
\begin{align}
 q^1 &= \operatorname{EM}( q^n)\\
q^1 &= \frac{3}{4}  q^n + \frac{1}{4}\operatorname{EM}( q^1)\\
 q^{n+1} &= \frac{1}{3} q^n + \frac{2}{3}\operatorname{EM}( q^1).
\end{align}
Converging to the Stratonovich equation with strong order $1/2$, and capturing higher order symmetric terms. For the monotone Euler-Maruyama scheme, we use established deterministic monotone methodology after redefining the combined velocity as in \cref{sec:minimising diffusion}. Namely, 
we let the left and right reconstructed cell face values be specified with a slope limiter
\begin{align}
q_i^{R} = q_i^n + 1/2 \psi(1/R^n_i)(q^n_{i+1} - q^n_{i}),\quad q_i^L = q_i^n - 1/2 \psi(R_i^n)(q_{i}^n-q_{i-1}^n),\quad R_i^n = (q_{i+1}^n-q_{i}^n)/(q_{i}^n-q_{i-1}^n).
\end{align}
We define a Euler Maruyama transport velocity $\hat{u}_{i+1/2}$, and use a upwind flux function
\begin{align}
F_{i+1/2}&=F(q^{R}_{i},q^{L}_{i+1},\hat{u}^n_{i+1/2}) = \hat{u}_{i+1/2}^{+}q^R_{i} + \hat{u}_{i+1/2}^{-}q^{L}_{i+1}, \quad \text{where}\quad \hat{u}_{i+1/2} = u_{i+1/2} + \Delta t^{-1}\sum_{p=1}^P \xi(x_{i+1/2})\Delta S^p.
\end{align}
Such that the flux-form Euler Maruyama numerical flow map, 
\begin{align}
q^{n+1}_i = \operatorname{EM} = q^n_i + \frac{\Delta t}{\Delta x
}(F_{i+1/2}-F_{i-1/2}),
\end{align} 
is a monotone function of flux contributing quadrature points $(q^R_i,q^R_{i-1},q^L_{i+1},q^L_{i})$. Conditional on the slope limiter, the flux contributing quadrature points can be locally bounded. Being a monotone function of locally bounded quadrature points is an important form of nonlinear stability (see \cite{zhang2010maximum,woodfield2024higher,zhang2011maximum,zhang2012maximum}), it implies sign preservation for compressible flow, as well as a notion of monotonicity weaker than Harten Hyman Lax Keyfitz \cite{harten1976finite} monotonicity. Furthermore, for incompressible velocities, the numerical method is provably TVD.
Sufficient conditions for a monotone solution on the slope limiter are $\psi(r) \in [0,\min(2,2r)]$ \cite{sweby1984high}, we employ the Koren limiter function 
\begin{align}
\psi(r) = \max(0,\min(1/3+2r/3,2r,2)
\end{align} given in \cite{koren1993robust}.  For comparison in this work, we also use the unlimited non-monotone-scheme when 
\begin{align}
\psi(r) = 1/3+2r/3.
\end{align}
In this case, the ensemble members will not be positivity preserving.

\subsection{Particle filter}\label{sec:particle filter}
This section aims at motivating the particle filter, introducing
the Bayesian filtering framework for a state space model, Sequential Importance Sampling, and resampling, in the interest of this paper being self-contained. The particle filter exposition here is certainly not novel and is considered the most basic available. For a more comprehensive guide see for example,  \cite{fearnhead2018particle,reich2015probabilistic,bain2009fundamentals,crisan2002survey,chopin2004central,gustafsson2010particle,reich2015probabilistic} and references therein.

For those wishing to skip this introduction and see the results in \cref{sec:Numerical demonstrations with data assimilation}, the method we employ is Sequential Importance Sampling (SIS) with resampling, a Sequential Monte Carlo method. More specifically we use the bootstrap particle filter of \cite{gordon1993novel}, however, we do not use multinomial resampling \cite{doucet2009tutorial} and instead use the systematic resampling of Kitagawa \cite{kitagawa1996monte}. We resample when the effective sample size falls below half of the number of ensemble members. \newline

In the Bayesian Filtering Framework for a state-space model, the goal is to estimate the hidden state $x_k$ at time $t_k$, given a sequence of noisy observations $y_{1: k}=\left\{y_1, y_2, \ldots, y_k\right\}$. The state transition model $
x_k \sim p(x_k |x_{k-1})
$ describes how the state evolves. The observation model $
y_k \sim p(y_k| x_k)
$ describes the likelihood of the observation given the state. We have changed the notation $q\mapsto x$ to align with the particle filtering literature. 
\newline
In a recursive theoretical setting, one supposes that one already has the posterior distribution from the previous time step $p(x_{k-1}|y_{1:k-1})$, and one has access to the transition probability distribution $p(x_k| x_{k-1})$ associated with the forward model advancing from state $x_{k-1}$, to $x_{k}$. The prior distribution $p(x_k|y_{1:k-1})$ could then be theoretically attained by integrating over all possible states for $x_{k-1}$ against the transition probability distribution $p(x_k|x_{k-1})$, as follows
\begin{align}
\underbrace{p(x_k | y_{1: k-1})}_{\text{prior}}=\int \underbrace{p(x_k|x_{k-1})}_{\text{transition}} \underbrace{p(x_{k-1} | y_{1: k-1})}_{\text{posterior at }k-1} d x_{k-1}. \label{eq:posterior evolution}
\end{align}
The state transition probability distribution $p(x_k|x_{k-1})$ accounts for uncertainty associated with the forward model. As one observes the data $y_{k}$, one wishes to update the prior distribution into a posterior distribution $p\left(x_k | y_{1: k}\right)$. This is achieved through the following recursive relationship for the posterior
\begin{align}
p(x_k | y_{1: k})=\frac{p(y_k | x_k)}{p(y_k | y_{1: k-1})} p(x_k | y_{1: k-1}).
\end{align}
This expression is well known and arises from Bayes Rule under the usual assumptions that the hidden state is Markov, and the observations are conditionally independent of the process. $p\left(y_k | y_{1: k-1}\right)$ is a normalizing constant. In practice, one cannot perform such an integration, or have access to a transition probability distribution. Instead Monte Carlo (Sequential) Importance Sampling is performed, particles $\lbrace x^{(e)} \rbrace_{e=1}^{E}$ are drawn from an easier to sample, proposal distribution, denoted $q$ and then associated with an importance weight $\lbrace w^{E}\rbrace_{e=1}^{E}$ according to a likelihood. The transition to the next step is typically accounted for with a forward model. \newline

A Monte-Carlo technique called Importance Sampling (IS) can be performed to estimate the properties of hard-to-evaluate distributions. 
The idea is to draw samples from an easier-to-sample proposal distribution and use importance weights to compensate. However in the non-recursive setting (in the absence of a resampling step) as data becomes available, one would have to recalculate all the importance weights. Instead, if the proposal distribution is assumed to be recursive, importance weights can be updated rather than recalculated, this is known as \emph{Sequential} Importance Sampling (SIS). Particle filters are a class of sequential Monte Carlo methods which approximate the posterior distribution $p\left(x_k | y_{1: k}\right)$ using a set of $E$ weighted Dirac delta functions $\left\lbrace x_k^{(e)}, w_k^{(e)}\right\rbrace_{e=1}^E$, as follows,
\begin{align}
p\left(x_k | y_{1: k}\right) \approx \sum_{e=1}^E w_k^{(e)} \delta\left(x_k-x_k^{(e)}\right).
\end{align}
Where $\delta$ denotes the Dirac delta function, and $w_k^{(e)}$ denotes the $e$-th importance weight at time $t_k$. Here $x_k^{(e)}$ denotes the $e$-th particle.
The importance weight of each particle in the recursive setting can be shown to be \cite{arulampalam2002tutorial} updated as follows 
\begin{align}   
w_k^{(e)} \propto w_{k-1}^{(e)} \frac{p\left(y_k | x_k^{(e)}\right)  p\left(x_k^{(e)} | x_{k-1}^{(e)}\right)}{q\left(x_k^{(e)} | x_{k-1}^{(e)}, y_k\right)} = w_{k-1}^{(e)} p\left(y_k | x_k^{(e)}\right),
\end{align}
where it is convenient to assume that the proposal distribution $q\left(x_k^{(e)} | x_{k-1}^{(e)}, y_k\right)$ used to generate new particles is itself the transition model, allowing a cancelling in the denominator and numerator. As remarked upon in \cite{arulampalam2002tutorial} it is convenient and simple to pick the importance density to be the prior in this manner but it is not the only option. The probability of observing $y_k$ given the state $x_k$ denoted $p\left(y_k | x_k\right)$ is assumed Gaussian so that 
\begin{align}
p\left(y_k | x_k^{(e)}\right)\propto \exp \left(-\frac{1}{2}\left(y_k-h\left(x_k^{(e)}\right)\right)^{T} R^{-1}\left(y_k-h\left(x_k^{(e)}\right)\right)\right), 
\end{align}
where $R$ is the covariance matrix of the observation noise and $h$ is the observation operator. Furthermore, we additionally assumed that $R$ is diagonal with $\sigma^2 I_{m\times m}$, where $\sigma>0$.
So the recalculation of the weights is done using the following formula
\begin{align}
w_{k}^{(e)} = w_{k-1}^{(e)} \exp{\left(-\frac{|| y_k - h(x_{k}^{(e)})||^2_2}{2\sigma^2}\right)}.\label{eq:weights_update}
\end{align}

A resampling step helps to mitigate particle degeneracy, where only a few particles have significant weights. A new set of particles $\left\{x_k^{(e)}\right\}_{e=1}^E$ is created by sampling from the current set but with replacement/duplication according to their likelihood weights $\left\{w_k^{(e)}\right\}_{e=1}^E$. As suggested in \cite{kong1994sequential} for computational efficiency reasons we resample when particle degeneracy starts to occur, we use the effective sample size diagnostic, given by \begin{align}
    N_{ess} =  \left(\sum_{e=1}^{E} w^{(e)}_{k}\right)^{-1}\label{eq:ess}.
\end{align}

Resampling is done when this falls below a threshold, we choose throughout this work $N_{ess}\leq E/2$. The \textbf{Systematic resampling} algorithm we use is described in \cite{kitagawa1996monte} and summarised below. 
\begin{enumerate}
    \item Compute vector $\b C$ whose $j$-th component is the cumulative sum of the weights $C_j = \sum_{e=1}^{j}w^{(e)}$, for all $j\in \lbrace 1,...,E\rbrace$. 
    \item Compute vector $\b u$, whose $j$-th component is $u_j= u + (j-1)/E$, $j\in\lbrace 1,...,E \rbrace$ so that $\b u$ is made up of equispaced points offset by $u\sim \mathcal{U}(0,1/E)$. Such that $u_j$ are evenly spaced points in $[0,1]$.
    \item To resample we require the indices of the particles corresponding to the positions of $\b u$. More specifically for all elements of $\b u$, we find the indexes of the first position in $\b C$ where each value from $\b u$ can be inserted without violating the sequential order of the cumulative sum vector $\b C$. In the sense that $i$ is returned such that $C_{i-1} < u_j \leq C_{i}$, $\forall j$. This determines which interval and corresponding particles should be resampled according to $\b u$.
    \item Reset weights $1/E$. 
\end{enumerate}

In summary, as observed data $h(x_k)$ is available we update the weights \cref{eq:weights_update}, and the ESS \cref{eq:ess}. If the ESS \cref{eq:ess} is less than $E/2$, we compute $\b C$, $\b u$, and then use the ``np.searchsorted" function in Python, to generate indices for the systematic resampling. Subsequently, all weights are renormalised to be $1/E$. \newline

\textbf{Interpretation:}
Conservation properties and monotonicity properties during resampling are understood in the following sense.
During resampling, one replaces a monotonic-conservative solution trajectory with another monotonic-conservative solution trajectory. The idea is that the proposed forward model has monotonic conservative properties.

\section{Numerical experiments with data assimilation}\label{sec:Numerical demonstrations with data assimilation}

\subsection{Example 2a: The Twin Experiment, One-dimensional slope limiters}\label{sec:Example 2a: The Twin Experiment}
In this example, we test to what extent SSP time integration in combination with monotonic solving strategies, can improve the forecast skill of a stochastic transport model when the reference data is a specific realisation of the same SPDE (solved monotonically). This is an idealised filtering problem. To do so we produce four ensembles, 
\begin{enumerate}
    \item SSP - with monotonic solving strategies - without a particle filter.
    \item  Without monotonic solving strategies - without a particle filter. 
    \item SSP - with monotonic solving strategies - with a particle filter.
    \item Without monotonic solving strategies - with a particle filter.
\end{enumerate}

\begin{figure}[H]
\centering
\begin{subfigure}[t]{0.495\textwidth}
\centering
\includegraphics[width=.95\textwidth]{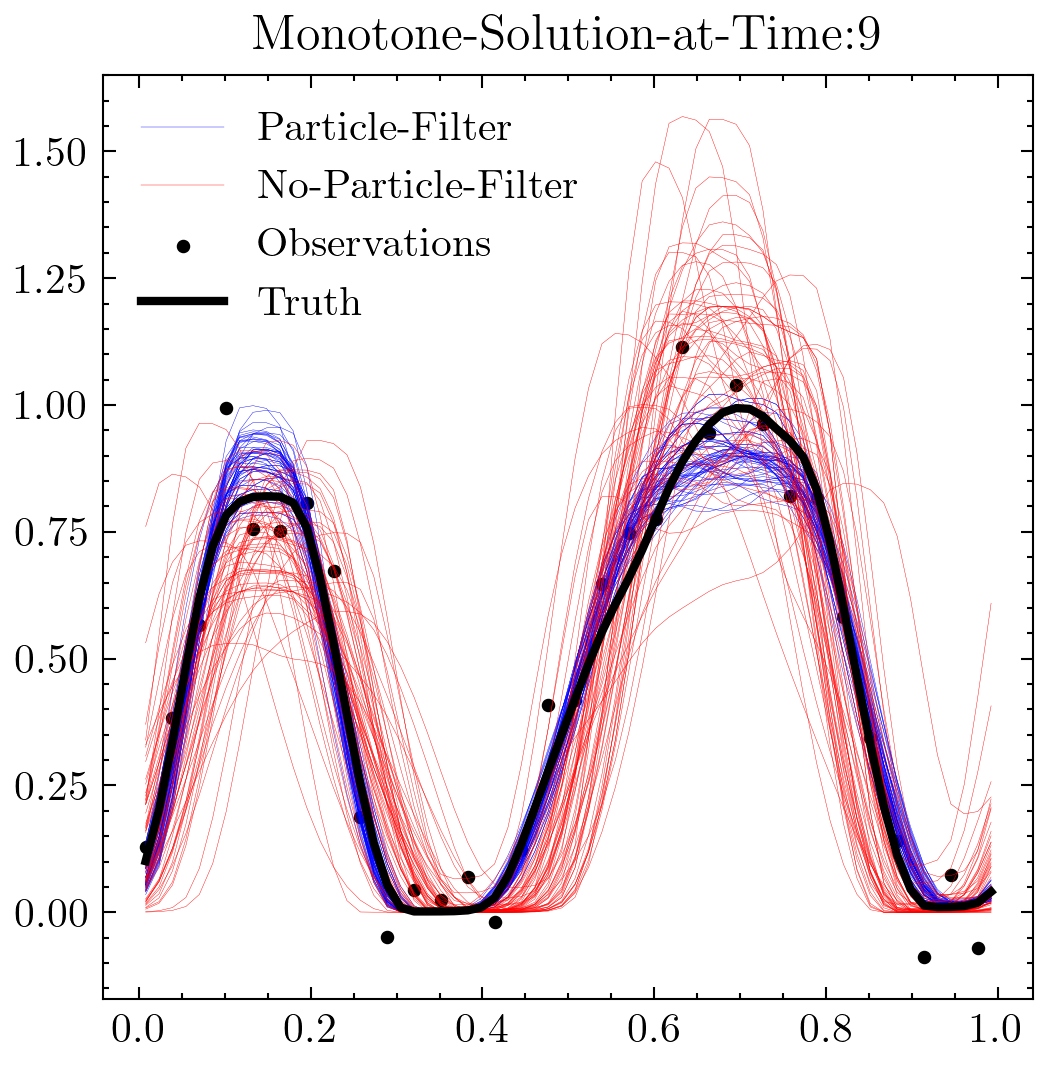}\caption{\hfill}\label{fig:TWIN_SSP_plot}
\end{subfigure}
\begin{subfigure}[t]{0.495\textwidth}
\centering
\includegraphics[width=.95\textwidth]{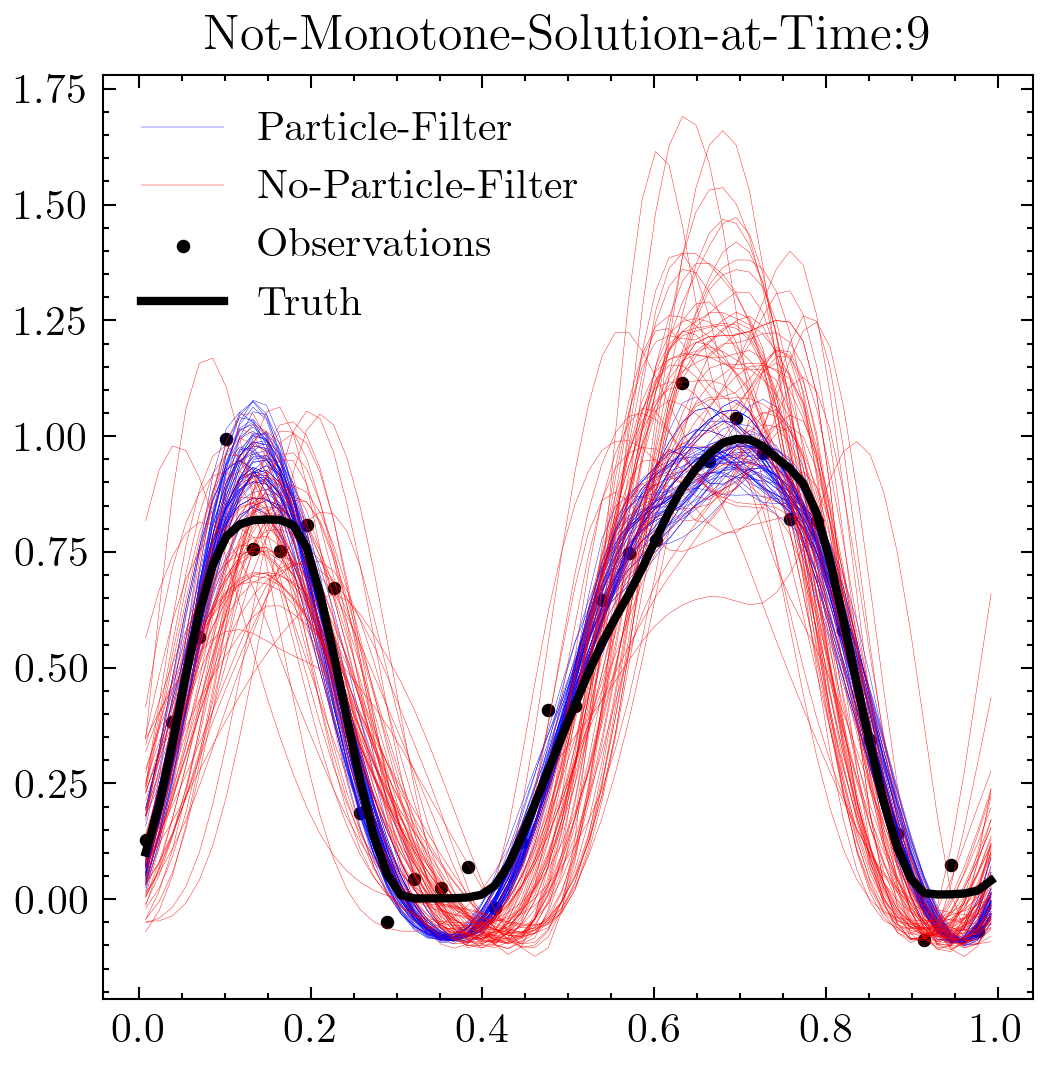}\caption{\hfill}\label{fig:TWIN_Not_SSP_plot}
\end{subfigure}
\caption{Ensemble solutions of \cref{eq:stochastic 1d transport problem}. In \cref{fig:TWIN_SSP_plot} monotonic solving strategies (SSP-limiting and bounded increments) are employed, in \cref{fig:TWIN_Not_SSP_plot} the limiting is not used. In black we plot the reference data a realisation of the SSP ensemble, and the noisy observations. In red we plot the ensemble solution without the particle filter, in blue we plot the ensemble solution with the particle filter.}
\end{figure}

We recover a reference data set $\lbrace q^{n}_{i,ref}\rbrace$ with the discrete property $q^{n}_{ref}\geq 0\implies q^{n+1}_{ref}\geq 0$, on the space time interval $(x,t) \in [0,1]\times[0,9]$, at resolution $n_x, n_{t}= 64,1024$. Plotted in \cref{fig:TWIN_SSP_plot}. We run a $E = 64$ member ensemble on the same space-time window at the same resolution starting from the same initial conditions using the same method and plotted in \cref{fig:TWIN_SSP_plot} with and without the particle filter. We also run another 64-member ensemble using the third-order unlimited scheme $\psi(r) = 1/3+2r/3$, with and without the particle filter, plotted in \cref{fig:TWIN_Not_SSP_plot}. 

Noisy observations occur every $n_{freq} = 16$ timesteps, on a $N_{obs}=32$ dimensional observation space. The measurement noise is drawn from a $32$-dimensional normal distribution at every observation window with standard deviation given by $\sigma = 0.1$. Weights are computed at every $n_{freq} = 16$ timesteps, as is the effective sample size $ESS = (\sum_{i}w_i^2)^{-1}$. Resampling (by a systematic resampling algorithm \cite{kitagawa1996monte}) occurs when the effective sample size, falls below $E/2$. This is described in \cref{sec:particle filter}.

In \cref{fig:TWIN_SSP_plot}, we plot the solution of the SSP slope-limited ensemble with a particle filter in blue, the ensemble without the particle filter is plotted in red. In \cref{fig:TWIN_SSP_plot} the reference data is denoted ``Truth" and plotted in black. In \cref{fig:TWIN_SSP_plot} noisy observations are also plotted.  In \cref{fig:TWIN_Not_SSP_plot} we plot the results of the same experiment, but for the unlimited scheme. In the absence of limiting as seen in \cref{fig:TWIN_Not_SSP_plot} unphysical oscillations and negative values occur and are not eliminated by using a particle filter.

In \cref{fig:TWIN_SSP_plot} and \cref{fig:TWIN_Not_SSP_plot} (independent of monotonic strategy)
the particle filter appears to track the truth by using the noisy measurement data, minimising the variance as compared to the ensembles without the filter. 
Ensemble members in \cref{fig:TWIN_SSP_plot}, remained positive, and had fewer undershoots and overshoots as compared with the non-monotonic method in \cref{fig:TWIN_Not_SSP_plot}. Whilst systematic resampling drastically improved the forecast capabilities of both ensembles, the systematic errors associated with a non-monotonic solving strategy were still present in \cref{fig:TWIN_Not_SSP_plot}. Motivating the potential necessity of a monotonic solving strategy.
\newline

\begin{figure}[H]
\centering
\begin{subfigure}[t]{0.495\textwidth}
\centering
\includegraphics[width=.95\textwidth]{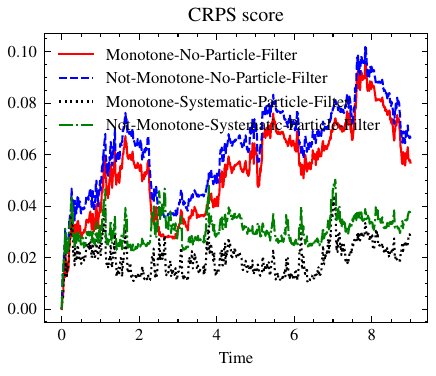}\caption{\hfill}\label{fig:TWIN_CRPS_score}
\end{subfigure}
\begin{subfigure}[t]{0.495\textwidth}
\centering
\includegraphics[width=.95\textwidth]{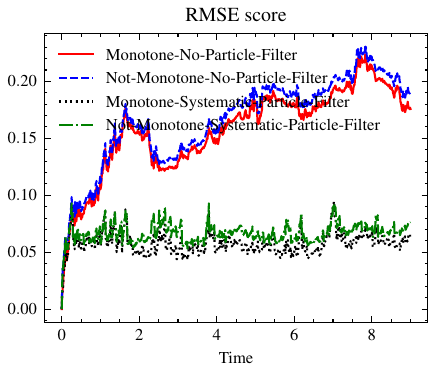}\caption{\hfill}\label{fig:TWIN_RMSE_score}
\end{subfigure}
\caption{In \cref{fig:TWIN_CRPS_score} and \cref{fig:TWIN_RMSE_score}, the Continuous Ranked Probability Score and Root Mean Square Error are plotted as a function of lead time respectively. The SSP slope limiting method is plotted with and without a particle filter (systematic resampling) in black and red respectively. The unlimited scheme is plotted with and without a particle filter in green and blue respectively. }
\end{figure}

We quantify some of the ensemble statistics associated with this experiment as a function of lead time. 
The CRPS is a proper score used to measure the difference between the ensemble and observation cumulative density functions, which is interpretable as integrating the Brier skill score over all possible thresholds. A CRPS of 0 indicates accuracy, whereas 1 indicates inaccuracy, for additional details regarding CRPS see \cite{arnold2023decompositions,gneiting2007strictly,gneiting2011comparing,hersbach2000decomposition,matheson1976scoring}. In \cref{fig:TWIN_CRPS_score} we plot the CRPS score for all 4 ensembles. In \cref{fig:TWIN_RMSE_score} we plot the RMSE score for all 4 ensembles. 

The CRPS score was highest in the Not-Monotone-No-Particle-Filter unlimited ensemble, second highest in the Monotone-No-Particle-Filter ensemble, third highest in the Not-Monotone-Systematic(resampled)-Particle-Filter ensemble and fourth highest (reaching the most accurate skill score) when using Monotone-Systematic-Particle-Filter ensemble. Indicating improved forecast skill from both the monotonic solving strategy and particle filtering, when the reference data is generated by the same monotonic forward model.

The use of a particle filter, allowed the ensembles to track the reference data and significantly decreased both RMSE and CRPS, for both SSP and Non-SSP solving strategies. Furthermore, with the particle filter the RMSE and CRPS show evidence of remaining bounded with lead time. Finally, we note that the CRPS score in \cref{fig:TWIN_CRPS_score} benefited from using both an SSP method with a monotonic solving strategy and a particle filter, producing better results than any other combination. This was also observed in the RMSE, but not so drastically. \newline

In this example, the improved forecast skill associated with the monotonic ensemble is expected as the reference data is generated from the monotonic (limited SSP) forward model. This experiment verifies that the particle filter is working, and is not a test indicating the utility of monotonic methodology. A different testing scenario is required to motivate the use of a monotonically solved ensemble, provided in the next section.

\subsection{Example 2b: Coarse-grained model reduction.}\label{sec:Example 2b: One-dimensional slope limiters}

In this example, we test to what extent SSP time integration in combination with monotonic solving strategies, can improve the forecast skill of a model when we define a reference solution, from a high-resolution simulation of the following PDE 
\begin{align}
\frac{d}{dt} q(t,x) + \frac{\partial }{\partial x}( u(t,x) q(t,x)) = 0, \quad q(0,x) = q_0(x),\quad x\in[0,1],
\label{eq:deterministic 1d transport problem}
\end{align}
a non-linear compressible advection equation with periodic boundary conditions. The velocity is the same compressible vector field as in \cref{eq:stochastic 1d transport problem}. We note that the reference data does not come from the ensemble forecast model primarily because it differs in both spatial and temporal resolution, as well as not having stochastic transport terms. This is a non-idealised data assimilation problem. We recover a high-resolution reference $\lbrace q^{n}_{i,ref}\rbrace$ with the discrete property $q^{n}_{ref}\geq 0\implies q^{n+1}_{ref}\geq 0$, using the previous numerical strategy. We use $n_x,n_t=256,4096$ for the high-resolution data. We have no reason to expect that the ensemble forecast model \cref{eq:stochastic 1d transport problem} is a reasonable proposal for the reference data generated by \cref{eq:deterministic 1d transport problem}, and test whether the particle filter can recover a forecast with skill. We hypothesise that because this reference data has an inherent monotonic property, the ensemble with a similar monotonic property will perform better than the ensemble without this property. 

\begin{figure}[H]
\centering
\begin{subfigure}[t]{0.495\textwidth}
\centering
\includegraphics[width=.95\textwidth]{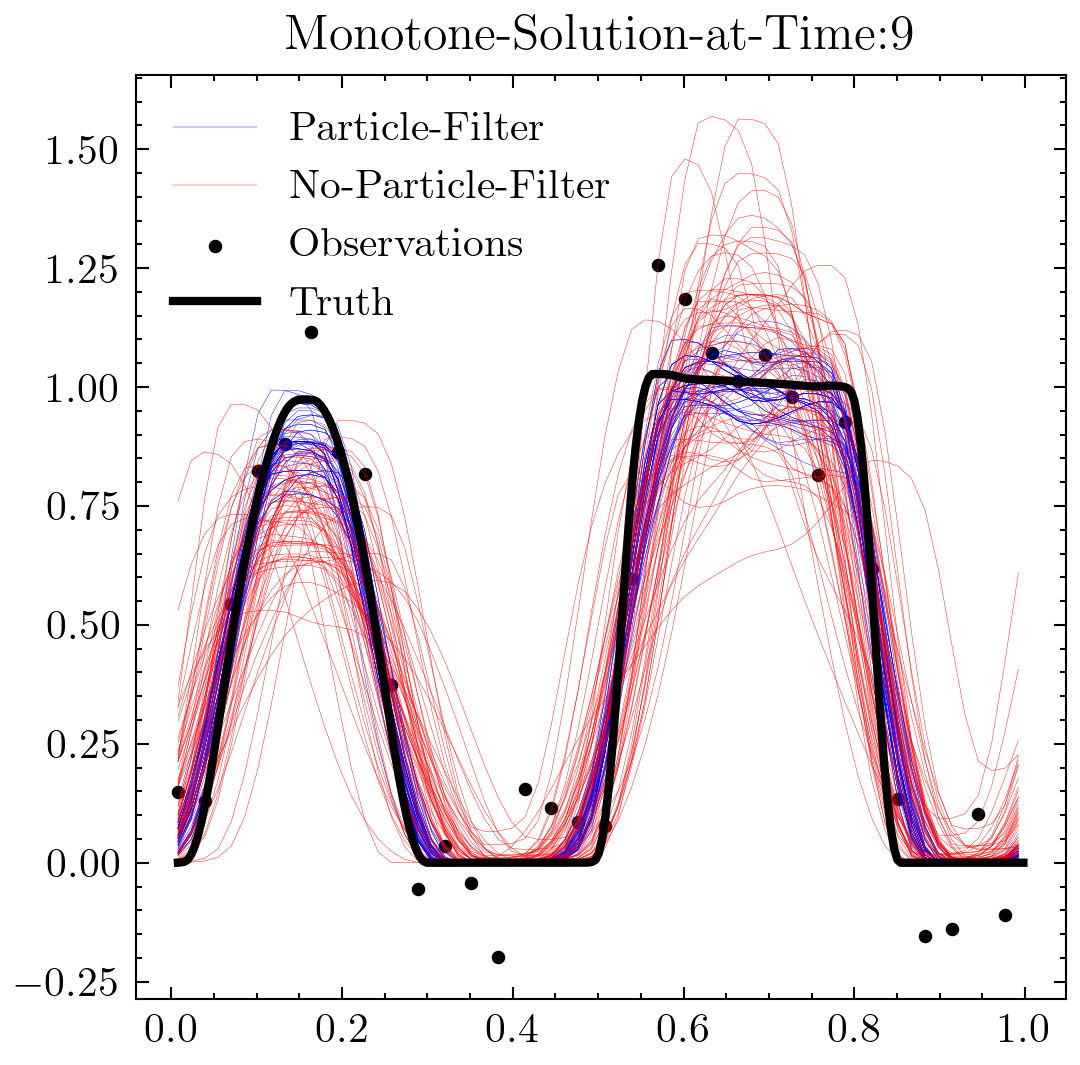}\caption{\hfill}\label{fig:SSP_plot}
\end{subfigure}
\begin{subfigure}[t]{0.495\textwidth}
\centering
\includegraphics[width=.95\textwidth]{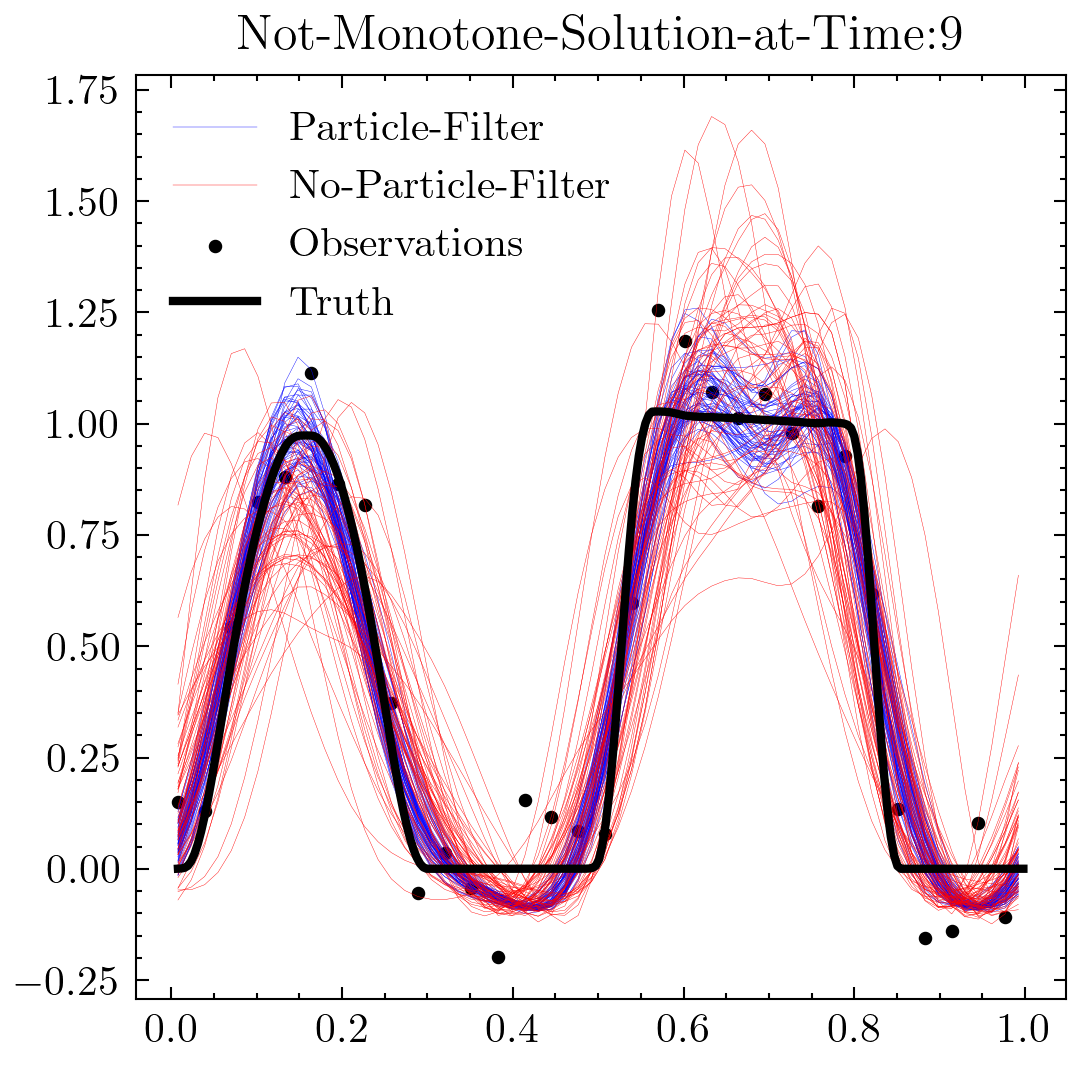}\caption{\hfill}\label{fig:Not_SSP_plot}
\end{subfigure}
\caption{Ensemble solutions of \cref{eq:stochastic 1d transport problem}. In \cref{fig:SSP_plot} monotonic solving strategies (SSP-limiting and bounded increments) are employed, in \cref{fig:Not_SSP_plot} the slope limiting is not used. In black we plot the reference data (the solution of a higher resolution PDE), and the noisy observations. In red we plot the ensemble solution without the particle filter, in blue we plot the ensemble solution with the particle filter.}
\end{figure}

The setup is the same as the previous example however the high-resolution reference data is subsampled further in both space and time to produce the same observation space (about one-fifth of one percent of available data). Observations can be monotonicity violating, due to the normally distributed observational noise with $\sigma=0.1$. The final time solution of all SSP slope limited ensembles is plotted in \cref{fig:SSP_plot}, and ensembles without limiting are plotted in \cref{fig:Not_SSP_plot}.

\begin{figure}[H]
\centering
\begin{subfigure}[t]{0.495\textwidth}
\centering
\includegraphics[width=.95\textwidth]{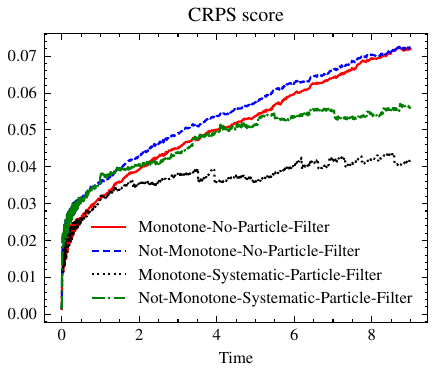}\caption{\hfill}\label{fig:Systematic_CRPS_score}
\end{subfigure}
\begin{subfigure}[t]{0.495\textwidth}
\centering
\includegraphics[width=.95\textwidth]{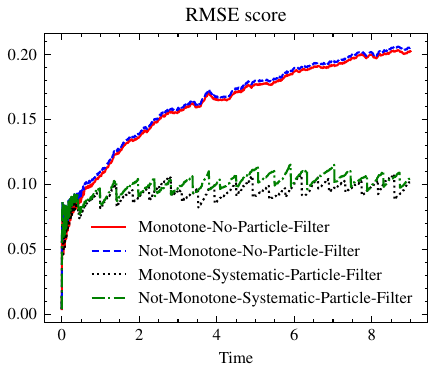}\caption{\hfill}\label{fig:Systematic_RMSE_score}
\end{subfigure}
\caption{In \cref{fig:Systematic_CRPS_score} and \cref{fig:Systematic_RMSE_score}, the Continuous Ranked Probability Score and Root Mean Square Error are plotted as a function of lead time respectively. The SSP slope limiting method is plotted with and without a particle filter (systematic resampling) in black and red respectively. The unlimited scheme is plotted with and without a particle filter in blue and green respectively. }
\end{figure}

Ensemble members in \cref{fig:SSP_plot}, remained positive, and had fewer undershoots and overshoots as compared with the non-monotonic method in \cref{fig:Not_SSP_plot}, both with and without the particle filter. In \cref{fig:Not_SSP_plot} the particle filter, drew samples from a proposal distribution effected by systematic undershoots and unphysical oscillations, arising from the numerical forward model. Whilst the systematic resampling particle filter drastically improved the forecast capabilities of both ensembles, the systematic errors associated with a non-monotonic solving strategy were still present in the ensemble in \cref{fig:Not_SSP_plot}. In \cref{fig:Systematic_CRPS_score} we plot the CRPS score for all 4 ensembles as a function of lead time, and in \cref{fig:Systematic_RMSE_score} we plot the RMSE as a function of lead time.

 In \cref{fig:Systematic_CRPS_score,fig:Systematic_RMSE_score} without the use of the particle filter both monotone and Non-monotone methods performed similarly in terms of CRPS and RMSE, with the monotone ensemble showing minor improvement over the non-monotone ensemble. The use of a particle filter, allowed the ensembles to track the reference data better and significantly decreased both RMSE and CRPS, for both SSP and Non SSP solving strategies \cref{fig:Systematic_CRPS_score,fig:Systematic_RMSE_score}. Furthermore, the RMSE shows evidence of remaining bounded with lead time (Long time plots confirm this for CRPS and RMSE available upon reasonable request). Finally, we note that the CRPS score in \cref{fig:Systematic_CRPS_score} benefited from using the combination of an SSP method monotonic solving strategy and a Particle Filter, producing better results than any other combination. This is also observed to a lesser extent in RMSE in \cref{fig:Systematic_RMSE_score}. We hypothesise that the increase in CRPS corresponded to the heuristic that an ensemble preserving physically motivated monotonic solutions can better represent latent data arising from a physical process with a similar monotonic property. 

Whilst the data does show an increase in CRPS when employing monotone strategies, the improvement is only modest in comparison to the decision to use data assimilation or not. Monotonic solutions will likely be more important in the context of coupled nonlinear equations, where the robustness gained by enforcing nonlinear stability and monotonic properties will be more important than the gain in the chosen skill score.


\section{Conclusion and outlook}

In \cref{sec:Example 2a: The Twin Experiment} and \cref{sec:Example 2b: One-dimensional slope limiters} we solve the one-dimensional compressible advection equation with compressible stochastic transport noise. This example tests to what extent SSP methodology (in combination with monotonic solving strategies) can be used to improve ensemble forecast skill both with and without data assimilation (in this case a particle filter). This was tested in an idealised setting when the data was from a realisation of the stochastic forward model and in a non-idealised setting when the data was the realisation of a higher resolution PDE with a monotone property. 

We conclude from our experiment and hypothesise more generally that it is possible to preserve nonlinear properties of SPDE's pathwise, and that this may be used in conjunction with a particle filter to produce ensembles whose members preserve discrete nonlinear properties. In the experiments here where the data also arise from a physical process with similar monotonic properties, this was shown to lead to improvements in CRPS and RMSE. However, this is likely problem specific and we do not advocate the use of monotonically produced ensembles for increasing raw RMSE or CRPS scores alone. We instead would like to advocate their use for systems for which monotonicity or positivity is deemed essential. 

More speculatively, establishing total variation estimates and monotonicity properties have been historically useful for establishing deterministic numerical methods converge to the unique weak entropy solution, for example by employing Kolmogorov compactness theorem. It may be possible to extend such results into the stochastic setting, for nonlinear equations under transport noise using the numerical methods presented here. The data assimilation procedure would then benefit from having ensemble members as convergent solutions of the proposed SPDE.

We also speculate that monotone methodology can be adapted for use in higher dimensional stochastic systems when the particle filter requires additional modification. 
In \cref{sec:degeneracy in the particle filter}, we discuss how one can jitter monotonically and conservatively, either as a re-realisation of the SPDE with correlated bounded increments or with a more cost-effective frozen-in-time jittering procedure. Experimentally, an example with tempering and jittering is shown to converge where the ordinary particle filter fails.


\section*{Acknowledgements}

JW is especially grateful to E. Fausti for discussions regarding jittering and tempering, as well as
A. Lobbe, E. Fausti, M. Kumar, D. Crisan, C. Cotter, and W. Pan for discussions regarding the particle filter, jittering, tempering and nudging. Thankful for the support and discussions with D. D. Holm, H. Weller,  O. Street, R. Hu, R. Wood leading to the improvement of this document. I would like to acknowledge the use of the SciencePlots library \cite{SciencePlots} by John D. Garrett. 
The corresponding author has been supported during the present work by the European Research Council (ERC) Synergy grant ``Stochastic Transport in Upper Ocean Dynamics" (STUOD) -- DLV-856408.

\bibliographystyle{abbrv}
\bibliography{refs.bib}

\appendix

\section{Degeneracy in the particle filter}\label{sec:degeneracy in the particle filter}
Unfortunately, the standard particle filter suffers from degeneracy, particularly in high dimensions. One approach to avoid degeneracy in the filter is the Merging particle filter proposed in \cite{nakano2007merging}, unfortunately, we remark the existing weight conditions of the merging particle filter are not compatible with a convex combination required for monotonicity in \cref{sec:merging particle filter}. Another approach to avoid degeneracy is the combination of jittering and tempering steps.

\subsection{Merging particle filter.}\label{sec:merging particle filter}
The merging particle filter, \cite{nakano2007merging} is a particle filter that attempts to avoid degeneracy, by making an $E\times T$ sized ensemble (with resampling) where $T\geq 3$. Then a weighted average is taken to produce a sized $E$ ensemble, 
\begin{align}
q^{(e)}_{**} = \sum_{t=1}^{T}  \alpha_t q_{*}^{(t,e)}. \label{eq:merging_pf_sum}
\end{align}
To preserve the mean and variance of the new filtered probability density function in the limit of infinite particles, the merging weights $\alpha_t$ are set to satisfy
the following conditions respectively
\begin{align}
\sum_{t=1}^T \alpha_t=1, \quad
\sum_{t=1}^T \alpha_t^2=1.
\label{eq:merging_pf_conditions}
\end{align}
These \cref{eq:merging_pf_conditions} conditions, imply that $\exists \alpha_t<0$, such that the representation \cref{eq:merging_pf_sum}, is not a convex combination. Preventing a natural type of monotonicity of the Merging particle filter. It would be interesting to relax the second condition in \cref{eq:merging_pf_conditions}, and sacrifice the variance of the filtered pdf in the limit of infinite particles, to ensure \cref{eq:merging_pf_sum} is a convex combination.

Jittering is a technique used to improve impoverishment associated with the standard particle filter. We demonstrate how jittering can be performed monotonically, by running the forward model monotonically with correlated bounded noise. Tempering is a technique used to improve the robustness of the particle filter by proposing several intermediate tempered likelihoods, requiring many jittering steps. Jittering is costly and one often uses additive noise, rather than running the forward model from the previous data assimilation stage. In between the efficiency of additive jittering and running the model again from the previous data assimilation stage. We show that one can perform frozen-in-time flux form monotonic jittering in combination with tempering. To preserve mass monotonic properties and not incur a huge computational cost associated with rerunning the entire model from the last time data was available. 

\subsection{Tempering}
Tempering aims to gradually change a proposal distribution into the posterior, by introducing intermediate ``tempered" distributions. This changes the particle filter in the following manner: 
\begin{method}[Tempering] We describe Tempering. 
\begin{enumerate}
    \item Define a sequence of tempering parameters 
$0=\beta_1<\beta_2<\cdots<\beta_{N_{temp}}=1
$.
\item For each $i$ in a sequence of tempering parameters $i \in \lbrace 0,...,N_{temp}-1\rbrace$:
\begin{enumerate}
\item 
We update the weights not using \cref{eq:weights_update}, but with the tempered likely-hood
\begin{align}
w^{(i+1,e)} = w^{(i,e)} \exp{\left(-\frac{|| y_k - h(x_{i+1}^{(e)})||^2_2 }{2\sigma^2}\underbrace{\left( \beta_{i+1}-\beta_{i} \right)}_{\text{temperature}}\right)}.
\end{align}
Such that weights are drawn $\propto\left[p\left(y_t \mid x_t^{(i)}\right)\right]^{\beta_{i+1}-\beta_{i}}
$, instead of $\propto\left[p\left(y_t \mid x_t^{(i)}\right)\right]^{1}
$, as in the usual particle filter.
\item We then normalise the weights to add to one,  and compute the ESS \cref{eq:ess}. 
\item When the ESS falls below a threshold (e.g $E/2$), Resample and reset weights to uniform. 
\item New proposals are generated for the next step in the tempering sequence. This can include a realisation of the transition model from the previous data assimilation stage (as in \cref{method:monotone jittering no tempering}). Or more cost-effective jittering methods such as a realisation of the transition model from the previous timestep. 
\end{enumerate}
\item Stop, output $x^{e}$
\end{enumerate}
\end{method}
Some advantages to such a procedure are; that resampling occurs many more times; and that weights are prevented from growing too large. The disadvantages to such a procedure are that many resampling steps can lead to degeneracy and the computational cost associated with creating new proposals can be non-trivial if running from the previous time data was available.
 
Jittering helps maintain particle diversity and can be designed to minimise the cost associated with creating proposals, so is often combined with tempering. We describe jittering in the absence of tempering below. 

\subsection{Monotone Jittering, forward model}

Jittering is a common approach for particle filters to avoid impoverishment, where many resampled particles are identical. In general, it is common to introduce additional additive noise to the ensemble members post-resampling and perform a recursive Metropolis-Hastings accept rejection algorithm. 

However, random additive jittering may lead to unphysical undershoots and overshoots and stops proposing a forecast based on the SPDE model alone (violating conservation laws and monotonic properties). Below, we describe one approach to introducing jittered particles that reruns the model forward solving the SPDE, in a monotonic mass-preserving manner from the last time data was available.
\begin{method}[Monotone Jittering]\label{method:monotone jittering no tempering} We describe a monotone jittering procedure as an SPDE model rerun from the last data assimilation time. This is in the absence of tempering and only one MCMC step (Metropolis-Hastings).
\begin{enumerate}
    \item Let $x^{n}$, denote the previous data assimilation step at $t^{n}$.
    \item We solve the SPDE up until the next time instance data is available $x^{n+n_{da}}$, at $t^{n} + \Delta t n_{da}$. With bounded increments $\Delta \widetilde{W}_1^{i,p}$ for $i=n,n+1,...,n+\Delta t n_{da}$, $p=1,...,P$. 
    \item At $t^{n} + \Delta t n_{da}$, data is available, and we perform the systematic resampled particle filter, as usual with output $x_{*,e}$.
    \item We generate a new uncorrelated bounded normal random variable $\Delta \widetilde{W}_2^{i,p}$, and create a correlated random variable with the parameter $r$ controlling the degree of correlation of the jittered particles \begin{align}
    \Delta \widetilde{W}_{3}^{i,p} = (r) \Delta \widetilde{W}_1^{i,p} + (1 - r)\Delta \widetilde{W}_2^{i,p}.
    \end{align}
    Importantly $r\in [0,1]$, and $\widetilde{W}_{1}^{i,p}\in [-A,A]$, $\widetilde{W}_{2}^{i,p}\in [-A,A]$ implies that $\widetilde{W}_{3}^{i,p}\in [-A,A]$.
    \item We go back to $x^n$, and rerun a new ensemble, using the correlated r.v $\Delta\widetilde{W}_{3}^{i,p}$ for new jittered particles $x_{**}$ at $t^{n} + \Delta t n_{da}$. Since the correlated r.v is bounded, one can attain the same monotonic property for new particles. 
    \item Metropolis-Hastings reject accept algorithm is performed based upon likelihoods.
    Let $u_e \sim U(0,1)$, $e=1,...,E$ be sampled from a uniform random variable. Then compute the likelihood of the observation given ensemble member for the jittered and un-jittered ensemble $l_{1,e} = p(y|x_{**,e})$
    $l_{2,e} = p(y|x_{*,e})$, for which we use the same probability as in \cref{eq:weights_update}. Where $u_e\leq \min(1,l_{1,e}/l_{2,e})$, we accept the new jittered proposal ensemble member 
    $x^{new,e} = x_{**,e}$, when  $u_e> \min(1,l_{1,e}/l_{2,e})$, we reject the jittered ensemble member and use
    $x^{new,e} = x_{*,e}$. This is performed for all ensemble members $e=1,...,E.$
    \item This is often performed multiple times with a stopping criterion. 
\end{enumerate}
\end{method}
 Some theoretical justification for jittering can be found in \cite{beskos2014stability}. More heuristically, jittering is commonly used to reintroduce diversity removed by the resampling algorithm (resampling typically introduces duplicate particles).  It is common to use the weights $\lbrace r,\sqrt{(1-r^2)} \rbrace$, which would require a different rescaling of the above argument, to run monotonically. It is also common, to rather than rerun the model, with a correlated random variable, to simply add Gaussian noise to the particles as to jitter them. This is much faster computationally, but violates the previously attained conservative form and monotone properties of the ensemble.

\subsection{Experiment one: Monotonic jittering, no tempering.}
\begin{figure}[H]
\centering
\begin{subfigure}[t]{0.295\textwidth}
\centering
\includegraphics[width=.95\textwidth]{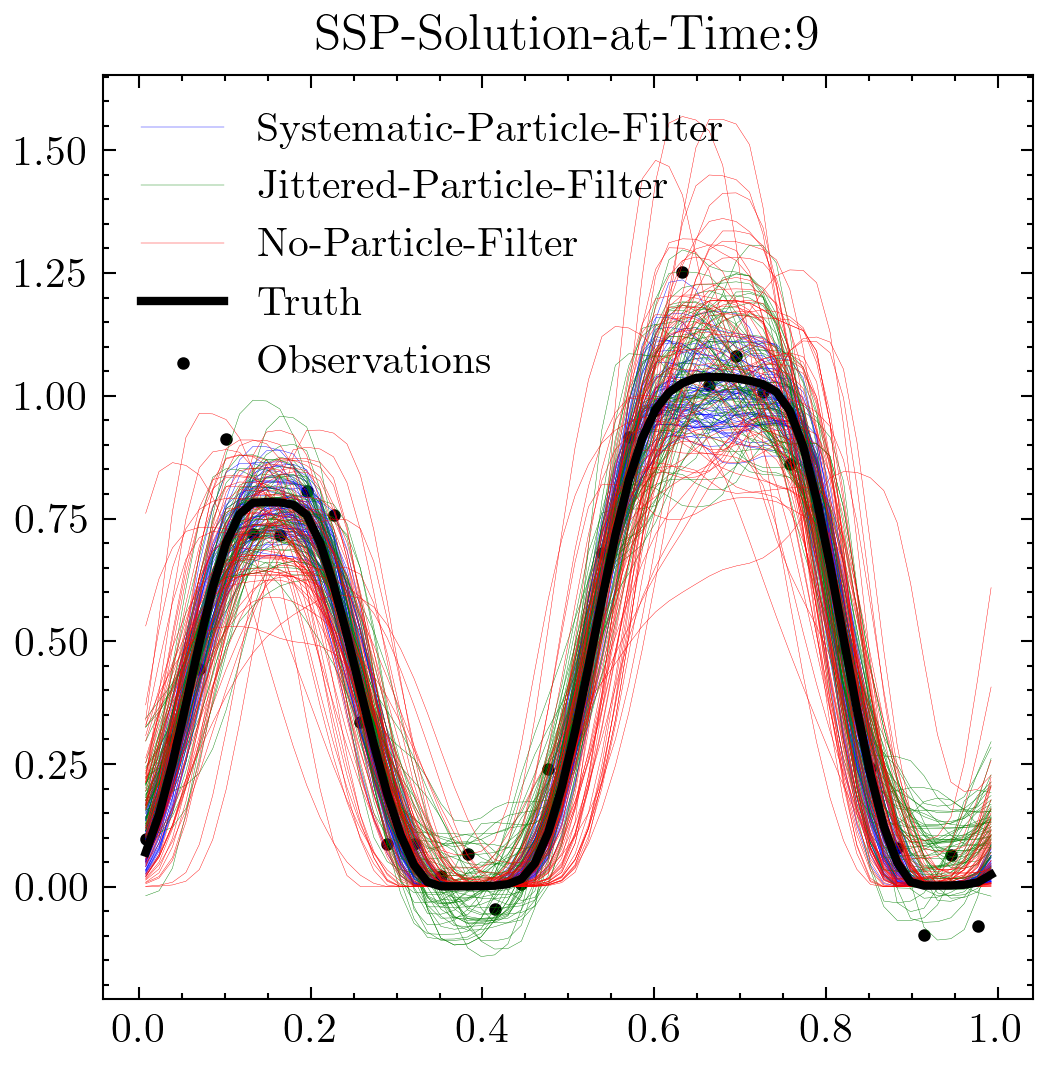}\caption{Additive Jittering\hfill}\label{fig:not-monotonic_jittering final time}
\end{subfigure}
\begin{subfigure}[t]{0.295\textwidth}
\centering
\includegraphics[width=.95\textwidth]{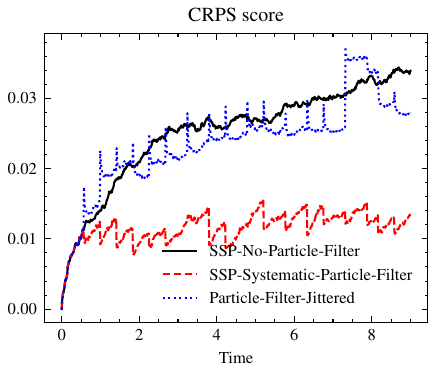}\caption{Additive Jittering\hfill}\label{fig:not-monotonic_jittering crps}
\end{subfigure}
\begin{subfigure}[t]{0.295\textwidth}
\centering
\includegraphics[width=.95\textwidth]{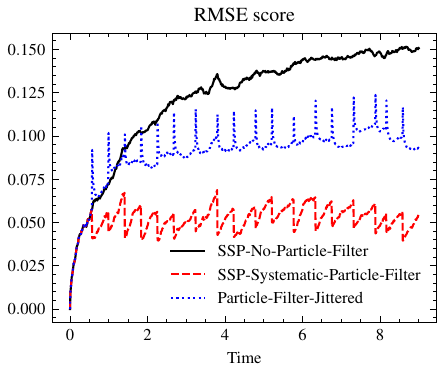}\caption{Additive Jittering\hfill}\label{fig:not-monotonic_jittering_rmse}
\end{subfigure}
\centering
\begin{subfigure}[t]{0.295\textwidth}
\centering
\includegraphics[width=.95\textwidth]{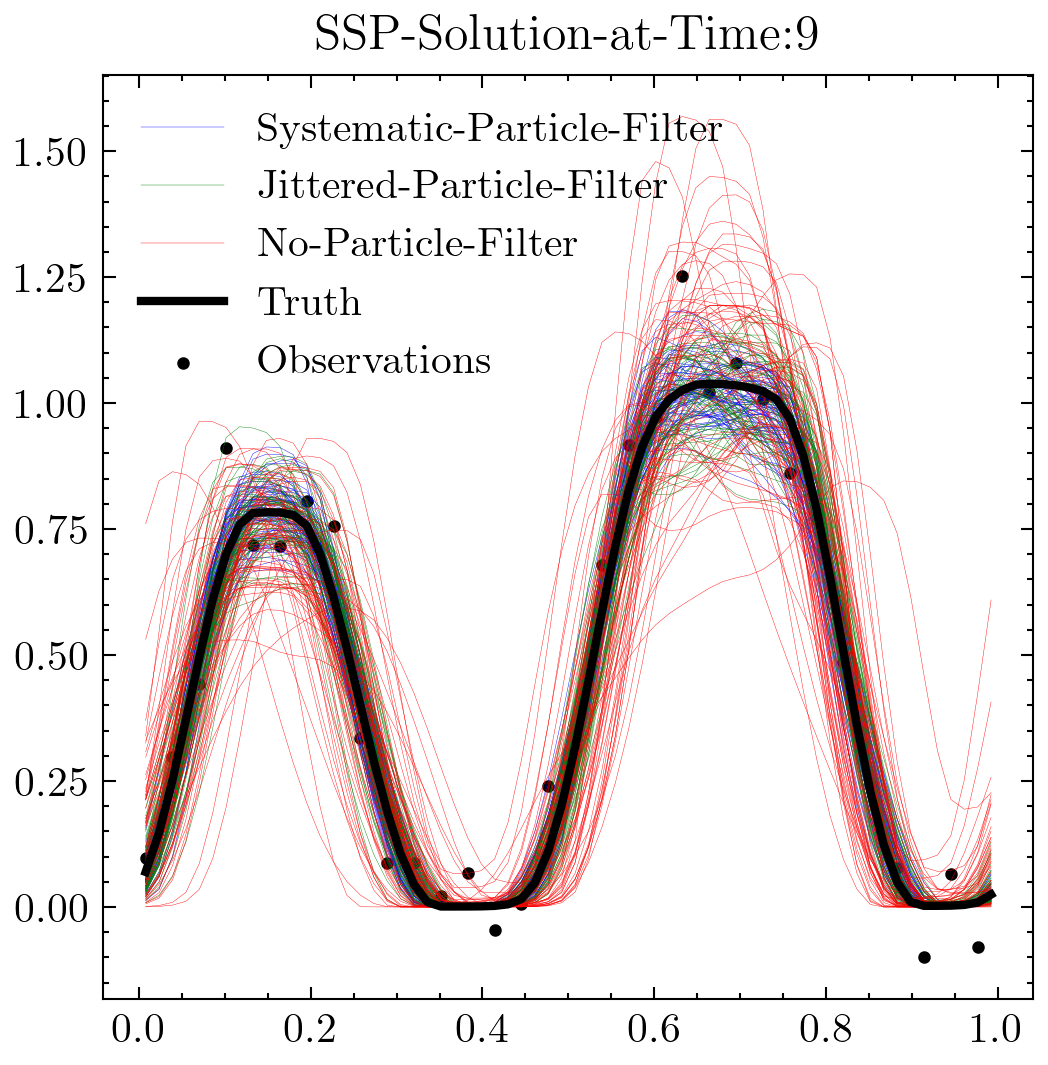}\caption{Monotonic Jittering\hfill}\label{fig:monotonic_jittering final time}
\end{subfigure}
\begin{subfigure}[t]{0.295\textwidth}
\centering
\includegraphics[width=.95\textwidth]{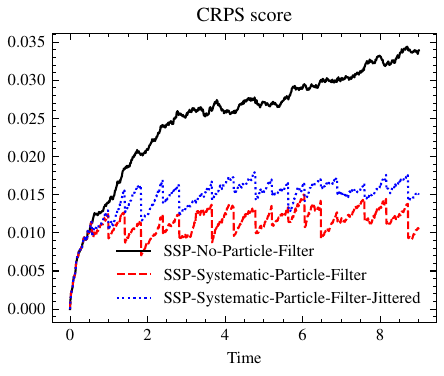}\caption{Monotonic Jittering\hfill}\label{fig:monotonic_jittering crps}
\end{subfigure}
\begin{subfigure}[t]{0.295\textwidth}
\centering
\includegraphics[width=.95\textwidth]{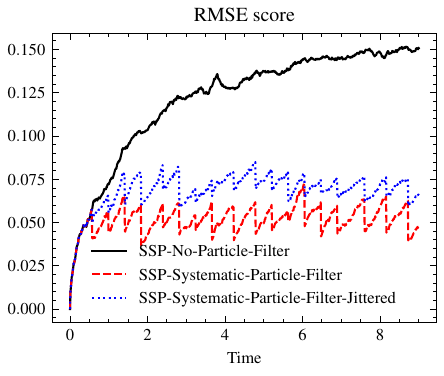}\caption{Monotonic Jittering\hfill}\label{fig:monotonic_jitteringRMSE}
\end{subfigure}
\caption{In row one, we use additive spacetime noise to jitter the particles, then perform a one step Metropolis-Hastings accept reject step. In row two, we propose ensemble members from rerunning the ensemble with correlated bounded random variables and perform a one step Metropolis-Hastings accept reject step.}
\label{fig:Monotonic one step Jitter}
\end{figure}

We perform the twin experiment, data is a realisation of the same SPDE as the ensemble. We use a one step Metropolis-Hastings accept-reject jittering (no tempering). We use the additive noise jittering and the monotonic jittering approach outlined in \cref{method:monotone jittering no tempering}. Both SPDE's are solved monotonically and conservatively, but additive jittering is not monotone or conservative. The results are plotted in \cref{fig:Monotonic one step Jitter}. Where in the first row we plot the additive noise jittering procedure's final timestep in \cref{fig:not-monotonic_jittering final time}, CRPS in \cref{fig:not-monotonic_jittering crps}, and RMSE in \cref{fig:not-monotonic_jittering_rmse}, compared to the bootstrap filter and no particle filter ensembles. In the second row we plot the same three plots but for the monotonically jittered ensembles; final timestep in \cref{fig:monotonic_jittering final time}, CRPS in \cref{fig:monotonic_jittering crps} and RMSE in \cref{fig:monotonic_jitteringRMSE}.  

In \cref{fig:not-monotonic_jittering final time} additive noise jittering violated positivity preservation. In \cref{fig:monotonic_jittering final time} we observe monotonic jittering \cref{method:monotone jittering no tempering} preserved positivity. In
\cref{fig:Monotonic one step Jitter}, there was not a significant enough gain in CRPS or RMSE to warrant the computational cost of 1-stage Metropolis-Hastings Jittering (in the absence of a tempering). We conclude it is possible to perform monotonic conservative jittering as a realisation of the SPDE solved monotonically, however, for this example jittering (monotonic or not) was not found to be useful. Jittering is well motivated when used in conjunction with tempering to prevent the particle filter from diverging, however, one may require a more cost-effective approach to monotonic jittering than \cref{method:monotone jittering no tempering} for sparse in-time observations, this will be introduced in the next section.

\subsection{Tempering and monotonic jittering}
We seek a sensible monotonic operator $\mathcal{M}:q^{n+1}\mapsto q^{n+1}$, such that dependent on a random variable, we can unconditionally preserve a desired norm or nonlinear stability property, and remain physically reasonable. Typical properties desired in weather-forecasting and for PDE conservation are
\begin{itemize}

    \item Mass conservation
    \item Positivity preservation
    \item Shape preservation
\end{itemize}
We would also like to not have to run the entire SPDE again from $q^n$ to $t^{n}+\Delta t n_{da}$ for different proposals. A sensible proposal could be to rerun the monotonic SPDE, from one step from $t^n + \Delta t n_{da} -1$ until $t^n + \Delta t n_{da}$.

\begin{method}[Monotonic Jittering and tempering]\label{method:monotonic jittering and tempering}We describe the combination of Jittering and temepering
\begin{enumerate}
    \item Let $q^{n}$, denote the previous data assimilation step at $t^{n}$.
    \item We solve the SPDE for the next time instance data is available $q^{n+n_{da}}$, at $t^{n} + \Delta t n_{da}$. with bounded increments $\Delta \widetilde{W}_1^{i,p}$ for $i=n,n+1,...,n+\Delta t n_{da}$, $p=1,...,P$. 
    \item At $t^{n} + \Delta t n_{da}$, data is available, and we perform the systematic resampled particle filter, as usual with output $q_{*,e}$, and weights $w^{(e)}$.
    \item \begin{enumerate}
    \item Define a sequence of tempering parameters 
$0=\beta_1<\beta_2<\cdots<\beta_{N_{temp}}=1
$, initialise the tempering step $x^{(e)}_0 = q_{*,e}, w^{(0,e)} = w^{(e)}$.
\item For each $i$ in a sequence of tempering parameters $i \in \lbrace 0,1,...,N_{temp}-1\rbrace$:
\begin{enumerate}
\item 
We update the weights with the tempered likelihood
\begin{align}
w^{(i+1,e)} = w^{(i,e)} \exp{\left(-\frac{|| y - h(x_{i+1}^{(e)})||^2_2 }{2\sigma^2}\left( \beta_{i+1}-\beta_{i} \right)\right)}.
\end{align}
\item We then normalise the weights,  and compute the ESS \cref{eq:ess}. 
\item When the ESS falls below a threshold (e.g $E/2$), Resample and reset weights to uniform. 
\item New proposals are generated for the next step in the tempering sequence. This is done frozen in time, using a monotonic jittering procedure:\label{item:frozen jitter}
\begin{enumerate}
    \item We generate new random variables, and a corresponding monotonic conservative operator $\mathcal{M}$.
    \item We apply the operator $\mathcal{M}:q_{*}\mapsto q_{**}$, to duplicated particles as to generate jittered particles. 
    \item (Return to 4(b)i.) with new $q_{i+1}$ jittered ensemble (until $i=N_{temp}-1$). 
    \end{enumerate}
\end{enumerate}
\item Stop, output assimilated ensemble.
\end{enumerate}
\end{enumerate}
\end{method}
Note that the MCMC Metropolis-Hastings accept-reject algorithm typically occurring in jittering alone is replaced by a resampling procedure where weights are prevented from becoming too large by tempering.  

The advantage to the jittering in \cref{method:monotonic jittering and tempering} over the jittering in \cref{method:monotone jittering no tempering} is one does not have to go back in time to rerun from $t^{n}$ until $t^n + \Delta t n_{da}$, to reproduce a monotonic ensemble. In this work we define a discrete Monotonic operator $\mathcal{M}_{\Delta t,\Delta x}$, as the \emph{numerical} solution to the following (no drift) nonlinear compressible transport equation, 
\begin{align}
d q(t,x) + \sum_{l=1}^{L}( \eta_{l}(x) q(t,x))_x \circ dW^{l}_t= 0, 
\end{align}
where we use the previous mass preserving monotonic method with bounded increments. The disadvantage and advantage to such a procedure is that the SPDE is not realised more than once to generate jittered ensemble proposals. Instead, a frozen-in-time conservative monotonic jittering procedure is performed to alleviate the cost associated with rerunning the forward model several times from the last time data was available (inherent in the tempering). It assumes sufficient representation and lack of bias in the stochastic monotone operator $\mathcal{M}$, in this case $\eta_l$.

\subsection{Experiment two: Tempering and Monotonic Jittering.}
We perform the coarse-grained model reduction experiment, where the reference data comes from a high-resolution PDE. However, in this example, we reduce the observation noise to $\sigma = 0.01$, such that the standard particle filter (systematic resampling bootstrap with resampling thresholding $N_{eff} = E/2$) fails to converge entirely.  

We then perform a Tempered and Jittered particle filter. We employ 100 instances of Tempering with linear tempering parameters with temperatures $\beta_{i+1}-\beta_i =  1/100$. We run two seperate Tempered-Jittered ensembles. The first uses space-time additive noise in the Jittering procedure, and the second uses the frozen-in-time monotonic flux form (mass preserving) drift-free advection jittering with $\eta_p = \xi_p$, $L=P$. 

\begin{figure}[H]
\centering
\begin{subfigure}[t]{0.295\textwidth}
\centering
\includegraphics[width=.95\textwidth]{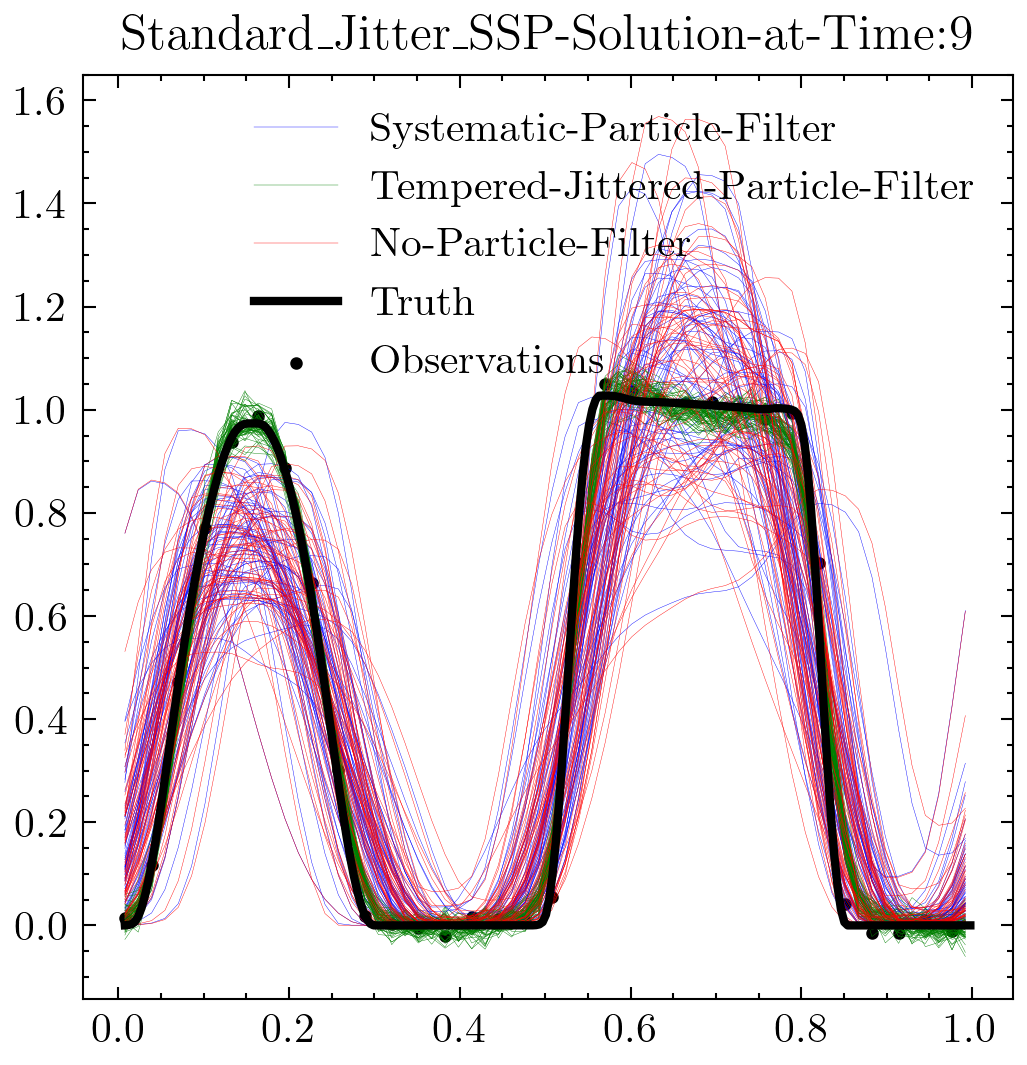}\caption{Additive jittering\hfill}\label{fig:Additive_Tempered_Jittered_Final_Time}
\end{subfigure}
\begin{subfigure}[t]{0.295\textwidth}
\centering
\includegraphics[width=.95\textwidth]{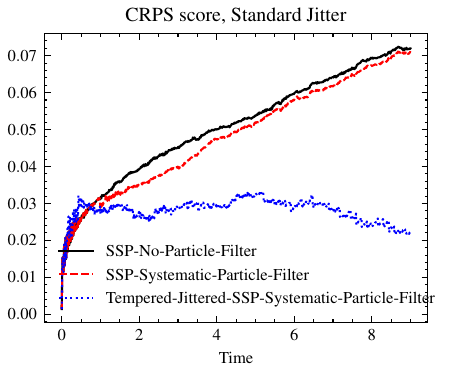}\caption{Additive jittering\hfill}\label{fig:Additive_Tempered_Jittered_CRPS}
\end{subfigure}
\centering
\begin{subfigure}[t]{0.295\textwidth}
\centering
\includegraphics[width=.95\textwidth]{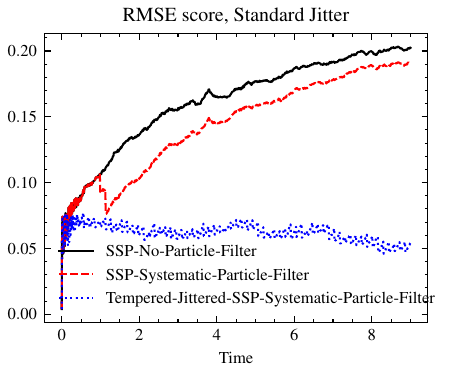}\caption{Additive jittering\hfill}\label{fig:Additive_Tempered_Jittered_RMSE}
\end{subfigure}\\
\begin{subfigure}[t]{0.295\textwidth}
\centering
\includegraphics[width=.95\textwidth]{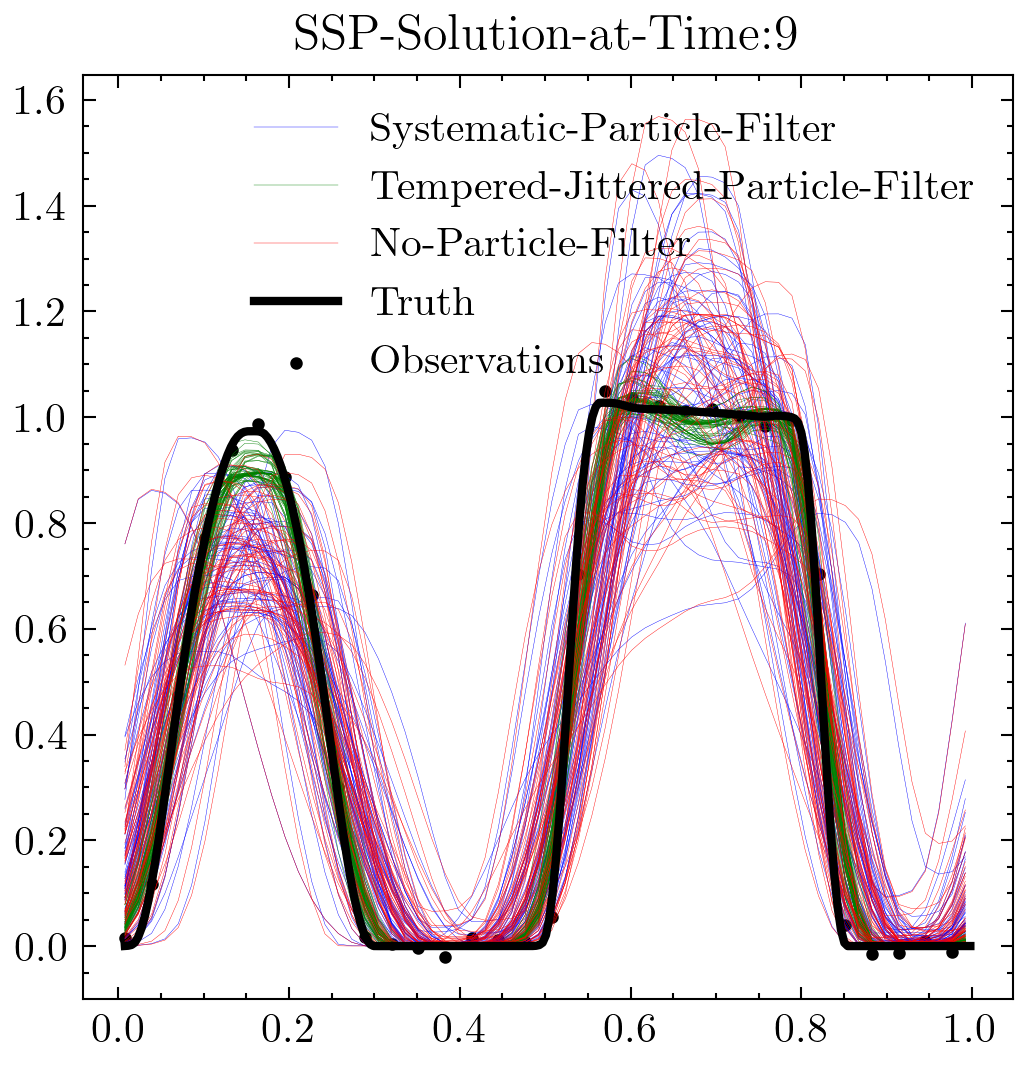}\caption{Monotonic jittering\hfill}\label{fig:MONOTONE_Tempered_Jittered_Final_Time}
\end{subfigure}
\begin{subfigure}[t]{0.295\textwidth}
\centering
\includegraphics[width=.95\textwidth]{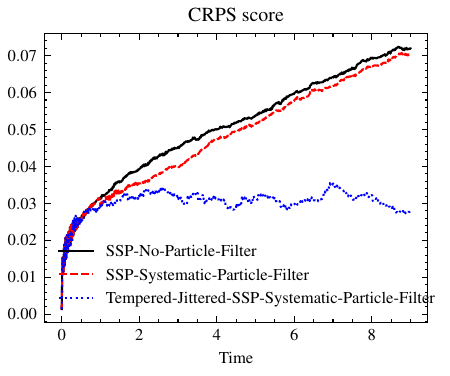}\caption{Monotonic jittering\hfill}\label{fig:MONOTONE_Tempered_Jittered_CRPS}
\end{subfigure}
\begin{subfigure}[t]{0.295\textwidth}
\centering
\includegraphics[width=.95\textwidth]{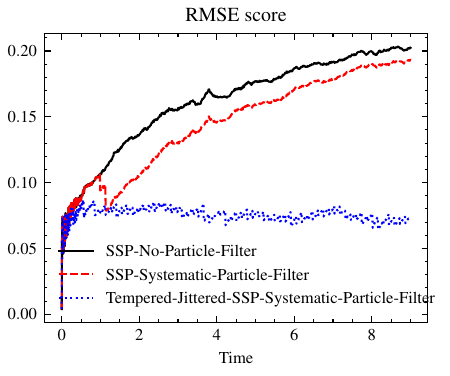}\caption{Monotonic jittering\hfill}\label{fig:MONOTONE_Tempered_Jittered_RMSE}
\end{subfigure}
\caption{Tempering and Jittering. }
\label{fig:Tempering and Monotonic Jittering}
\end{figure}

In \cref{fig:Additive_Tempered_Jittered_Final_Time}, we plot the final timestep of the unlimited ensemble in red, the standard particle filter ensemble in blue, the additive noise jittered ensemble in green. In \cref{fig:Additive_Tempered_Jittered_CRPS} and \cref{fig:Additive_Tempered_Jittered_RMSE} we plot the CRPS and the RMSE of the same three ensembles, the unlimited ensemble is plotted in black, the standard particle filter is plotted in red, and the additive noise jittered is plotted in blue. 
In \cref{fig:MONOTONE_Tempered_Jittered_Final_Time}, we plot the final timestep of the unlimited (red), standard particle filter (blue), monotonically jittered solution (green). In \cref{fig:MONOTONE_Tempered_Jittered_CRPS} and \cref{fig:MONOTONE_Tempered_Jittered_RMSE} we plot the CRPS and the RMSE of the same three ensembles, unlimited ensemble is plotted in black, the standard particle filter in red, and the monotonically jittered in blue.

When comparing CRPS and RMSE in \cref{fig:MONOTONE_Tempered_Jittered_CRPS,fig:Additive_Tempered_Jittered_CRPS,fig:MONOTONE_Tempered_Jittered_RMSE,fig:Additive_Tempered_Jittered_RMSE}, we observe that the standard particle filter does not stabilise in terms of skill score. We also observe that the tempered jittered particle filter stabilises in terms of skill score. The additive and monotonic approaches to jittering perform similarly regarding CRPS and RMSE. In \cref{fig:Additive_Tempered_Jittered_Final_Time} the additive noise destroys the positivity preservation, in \cref{fig:MONOTONE_Tempered_Jittered_Final_Time} the monotonically jittered ensemble preserves positivity. In \cref{fig:Additive_Tempered_Jittered_Final_Time}, the filtered additively jittered particle filter proposes an ensemble that tracks the truth, despite the violation of the conservation of mass, and positivity. In \cref{fig:MONOTONE_Tempered_Jittered_Final_Time}, the filtered monotonic flux form jittered particle filter proposes an ensemble that tracks the truth, whilst preserving mass and preserving positivity. 

One can conclude that it is possible to jitter monotonically and conservatively, allowing a tempered particle filter to converge when the standard particle filter diverges. However, one should be aware that this procedure may inherit bias from the specific monotonic operator chosen, such as the slight clipping of extrema present in \cref{fig:MONOTONE_Tempered_Jittered_Final_Time}, but not in \cref{fig:Additive_Tempered_Jittered_Final_Time}. We have deliberately chosen compressible vector fields in the stochastic basis for the noise, this was to allow the solution to grow and shrink locally under mass-preserving monotonic transport, the choice of stochastic basis will likely play a large part in the success of the filter and should be calibrated specifically to the data or the probability law of the SPDE/SDE which one wishes to model, rather than the arbitrary choice made in this work, some model aware approached are outlined in \cite{woodfield2024stochastic,leahy2024scaled}. Furthermore, it is tempering that overcomes the particular degeneracy problem here, jittering will be more

\end{document}